\documentclass[aps,prb,twocolumn,showpacs]{revtex4-1}

\usepackage{amsmath}
\usepackage{color}
\usepackage{amssymb}

\usepackage{graphicx}

\begin{document}

\title{Magnetostatics and optics of noncentrosymmetric metals}

\author{V.P.Mineev}
\affiliation{CEA / INAC /SPSMS, 38054 Grenoble, France}

\begin{abstract}

The current in noncentrosymmetric metals in normal and superconducting state is found in
frame of linear response theory. In line with usual terms corresponding diamagnetic response, 
the Landau diamagnetism and the Pauli paramagnetism the general expression contains also 
the terms corresponding to spacial dispersion specific for a medium without space parity. This, so 
called, gyrotropic current is calculated in zero frequency case as well in infrared frequency region. 
The static gyrotropic current yields negligibly small correction to the London magnetostatics in 
a superconductor without inversion center. Whereas the high frequency response produces the natural-optical 
activity revealed f.e. in the Kerr effect. The magnitude of the Kerr angle in
infrared frequency region is proved to be in reasonable correspondence with recently reported
observations of the Kerr effect in the pseudogap phase in several different high Tc materials.

\end{abstract}
\pacs{78.20.Ek, 74.25.N-, 74.20.Fg, 74.72.-h}

\date{\today}
\maketitle
\section{Introduction}
The normal state of high T$_c$ superconducting materials possesses many peculiar properties. 
The strange depletion of the density of states at the Fermi energy arising below some temperature T$^*$ 
 in the underdoped region of the phase diagram is one of them. It is probably  related to the onset of some nonsuperconducting electronic ordering.
 Indeed, quite recently several observations have been reported  \cite{Chang,Ashkar,Ghiringhelly} of  the short-range charge density wave ordering  detected by x-ray diffraction  arising at about the same temperature. Another peculiar property  is the  Kerr rotation in the reflected light polarization
 revealed practically in all families of the hole-doped cuprates: underdoped YBCO-123, \cite{Xia}
Hg-1201,\cite{Karapetyan} optimally doped Bi-2201, \cite{He} and 1/8 doped LBCO \cite{Kar}.
The Kerr onset temperature T$_K$ is somewhere in
the pseudogap regime. 
 This  implies a symmetry breaking
phase transition at T$_K$, despite the lack of 
thermodynamic evidence of such a transition (see, however, the recent paper by A. Shekhter et al \cite{Shekhter}).
 
 There are two peculiar features of the observed Kerr effects in cuprate materials. First, if
  the sample is cooled down through T$_K$ in  a field $H\pm60$ Oe,  and then measured in a zero field warm up,
  the  sign and value of the Kerr signal  is unchanged. Secondly,  recent measurements have demonstrated that the sign of the Kerr angle is the same for reflection on the opposite crystal surfaces.\cite{Karapetyan} Both of these observations are a strong indication that here we deal with the Kerr effect not due to ferromagnetism breaking the  time inversion symmetry but with the Kerr rotation of gyrotropic origin revealing itself in  media
 with broken space inversion symmetry. \cite{electr} The latter  property, called natural optical activity, is realized in a material with noncentrosymmetric crystal structure.
Metals without inversion symmetry have recently become a subject of considerable interest  following the  discovery of superconductivity in
CePt$_3$Si,
\cite{Bauer04}
UIr \cite{Akazawa04}, CeRhSi$_3$ 
\cite{Kimura05}, CeIrSi$_3$ \cite{Sugitani06},
Y$_2$C$_3$ \cite{Amano04},
Li$_2$(Pd$_{1-x}$,Pt$_x$)$_3$B \cite{LiPt-PdB},
KOs$_2$O$_6$ \cite{KOsO}, 
and other compounds.

The Kerr effect for light reflection from  media with chiral charge ordering of spinless electrons has been considered in the recent paper by P. Hosur and co-authors \cite{Hosur}. The natural optical activity in this type of model originates from
 specific terms in periodic crystal field and the corresponding modification of the electron energy spectrum of the noncentrosymmetric crystal. In the case of  slow enough noncentrosymmetric distortion,  it can be mathematically described as a gauge field (Berry curvature) effectively acting as an magnetic field.\cite{Orenstein}
 
On the other hand,  in  noncentrosymmetric media there is another mechanism for the natural optical activity related with Bychkov-Rashba type spin-orbital interaction. Such a theory has been developed by the author and Y. Yoshioka \cite{Mineev}.
In that  paper  calculations were  done at frequencies not exceeding the spin-orbital band splitting,  whereas
the Kerr effect measurements  on high T$_c$ materials \cite{Xia,Karapetyan,He,Kar}
have been performed in the infrared frequency region (more exactly at a wave length 1550 nm). So, it is appropriate to develop the corresponding theory for this frequency region. It is done in the present paper. 
 
The modifications of  electrodynamics introduced by space parity violation can be conveniently understood by starting from the expression for the displacement current
\begin{equation}
{\bf j}_d=\frac{\varepsilon}{4\pi}\frac{\partial{\bf E}}{\partial t}.
\end{equation}
The current changes sign ${\bf j}_d\to-{\bf j}_d$ both at time $
t\to -t$, and space 
${\bf r}\to -{\bf r}$  inversion. The latter property is lost in noncentrosymmetric media, hence the current expression has to be supplemented with the following terms \cite{Mineev}
\begin{equation}
{\bf j}_d\to{\bf j}_d+{\bf j}_g,~~~~~
{\bf j}_g=\lambda~\text{rot}~{\bf E}+\nu{\bf B},
\label{j_g}
\end{equation}
where ${\bf j}_g$ is so called gyrotropy current changing sign at time inversion
$
t\to -t$, ${\bf j}_g\to-{\bf j}_g$ but not at space inversion
${\bf r}\to -{\bf r}$, ${\bf j}_g\to{\bf j}_g$.
To guarantee these properties $\lambda$ must be an odd function of the time derivative $\partial/\partial t$ (or frequency),
$\nu$ is an even function of frequency. Owing the relation between the Fourier components of magnetic induction and  electric field $B_i=c e_{ijk}\frac{q_j}{\omega}E_k$, which is valid at finite frequency $\omega$ and wave vector ${\bf q}$ , the $\nu$ frequency  dependence can be included in the frequency dependence of $\lambda$.  Hence, $\nu$
can be taken as frequency independent.  Thus, the Fourier component of the frequency dependent part of the gyrotropy current can be written as
\begin{equation}
j_{gi}(\omega, {\bf q})=ie_{ijl}\lambda(\omega)q_jE_l(\omega, {\bf q}).
\label{gg}
\end{equation}

Eqs. (\ref{j_g}), (\ref{gg}) are valid in an isotropic gyrotropy medium,  otherwise the functions $\lambda$ and $\nu$ are  tensors, such that in general the dielectric permeability is 
\begin{equation}
\varepsilon_{ij}(\omega, {\bf q})=\varepsilon_{ij}(\omega, 0)+i\gamma_{ijl}(\omega)q_l,~~~~\gamma_{ijl}(\omega)=-\frac{4\pi i}{\omega}\lambda_{ijl}
(\omega),
\end{equation}
where the off-diagonal tensor $\gamma_{ijl}(\omega)=-\gamma_{jil}(\omega)$   is an even real function of frequency.\cite{electr} 
Like the Hall conductivity  in media with broken time inversion symmetry,  it describes  optical activity in the case of broken space inversion symmetry.

In addition to the usual electromagnetic field action\\ $S=\frac{1}{8\pi}\int dtd^3{\bf r}(\varepsilon{\bf E}^2-{\bf B}^2/\mu)$, 
the action in an isotropic gyrotropic medium contains a specific term
\begin{equation}
S_g=-\frac{1}{2c}\int dtd^3{\bf r}\left\{{\bf B} \lambda{\bf E}+{\bf B}\nu\left[{\bf A}-\frac{\hbar c}{2e}\nabla\varphi\right]\right\}.
\end{equation}
The gyrotropy current and magnetic moment are obtained from here as the variational derivatives\cite{Mineev}
\begin{equation}
{\bf j}_g=-c\frac{\delta S_g}{\delta{\bf A}}=\lambda~\text{rot}~{\bf E}+\nu{\bf B},
\end{equation}
\begin{equation}
{\bf M}_g=-\frac{\delta S_g}{\delta{\bf B}}=\frac{1}{2c}\lambda{\bf E}+\frac{1}{2c}\nu\left[{\bf A}-\frac{\hbar c}{2e}\nabla\varphi\right].
\end{equation}
Owing to gauge invariance, the last term $\propto \nu$ in all these expressions  exists only in the superconducting state.

We shall use
the current response to the electromagnetic field with finite frequency and wave vector. This allows us to discuss   magnetostatic phenomena in  noncentrosymmetric superconductors 
 like  field penetration, magnetoelectric effects and also 
  magnetic susceptibility anisotropy at finite wave vector. After that, we consider the optical properties of noncentrosymmetric metals. 
 For simplicity, we shall discuss mostly the isotropic medium situation.
 
 Finally, we will compare our results originating from the spin-orbital coupling in noncentrosymmetric materials with predictions of  the model operating with chiral charge ordering of spinless electrons \cite{Hosur}. 

\section{Current}
The electron spectrum of a noncentrosymmetric metal has the following form \cite{Sam07,SamMin08}
\begin{equation}
\xi_{\alpha\beta}({\bf k})=(\varepsilon({\bf k})-\mu)\delta_{\alpha\beta}+
\mbox{\boldmath$\gamma$}({\bf k})\mbox{\boldmath$\sigma$}_{\alpha\beta}-\mu_B{\bf B}\mbox{\boldmath$\sigma$}_{\alpha\beta},
\label{spectrum}
\end{equation}
where $\mu_B$ is the Bohr magneton, and the spin-orbital coupling  is
 determined by the dot product of the Pauli matrix vector $\mbox{\boldmath$\sigma$}=(\sigma_x,\sigma_y,\sigma_z)$ and  pseudovector $\mbox{\boldmath$\gamma$}({\bf k})$, which is  odd with respect to momentum
 $\mbox{\boldmath$\gamma$}(-{\bf k})=-$\mbox{\boldmath$\gamma$}({\bf k})$ $  and specific to each noncentrosymmetric crystal structure. 
For the cubic group $G=O$,
which describes the point symmetry of
Li$_2$(Pd$_{1-x}$,Pt$_x$)$_3$B, as well in isotropic media, the simplest form compatible with
the symmetry requirements is
\begin{equation}
\label{gamma_O}
 \mbox{\boldmath$\gamma$}({\bf k})=\gamma{\bf k},
\end{equation}
where $\gamma$ is a constant. 
For the tetragonal group
$G={C}_{4v}$, which is relevant for CePt$_3$Si,
CeRhSi$_3$ and CeIrSi$_3$, the spin-orbit coupling is given by
\begin{equation}
\label{gamma C4v}
\mbox{\boldmath$\gamma$}({\bf k})=\gamma_\perp(k_y\hat x-k_x\hat y)
    +\gamma_\parallel k_xk_yk_z(k_x^2-k_y^2)\hat z.
\end{equation}

The current response of a clean metal to the electromagnetic field at finite ${\bf q}$ and $\omega$ is\cite{fn}
\begin{widetext}
\begin{equation}
\label{basic}
 j_i(\omega_n,{\bf q})= -\frac{e^2}{c}Tr\left [\hat m^{-1}_{ij}\hat n_e +
\int  \frac{d^3k}{(2\pi)^3}\text{T}\sum_{m=-\infty}^{\infty} 
\{  \hat v_i({\bf k})\hat G^{(0)}(K_+)\hat v_j({\bf k})\hat G^{(0)}(K_-)
- \hat v_i({\bf k})\hat F^{(0)}(K_+)\hat v_j^t(-{\bf k})\hat F^{+(0)}(K_-)\}
\right ]{A_j}(\omega_n,{\bf q}).
\end{equation}
\end{widetext}
Throughout the article, we put   $\hbar=1$.
All the quantities here such as single electron energy 
$
\xi_{\alpha\beta}({\bf k}), $
velocity
\begin{equation}
{\bf v}_{\alpha\beta}({\bf k})=-(c/e)\frac{\partial\xi_{\alpha\beta}({\bf k-e{\bf A}/c})}{\partial{\bf A}}({{\bf A}\to0}),
\label{v}
\end{equation}
the inverse effective mass
$(m^{-1}_{ij})_{\alpha\beta}=\partial^2\xi_{\alpha\beta}({\bf k})/\partial k_i\partial k_j$, the Green functions 
$G_{\alpha\beta}(\tau_1,{\bf k};\tau_2,{\bf k}')=-\langle T_\tau a_{{\bf k}\alpha}(\tau_1)a^\dagger_{{\bf k}'\beta}(\tau_2)\rangle$ and $F_{\alpha\beta}(\tau_1,{\bf k};\tau_2,{\bf k}')=\langle T_\tau a_{{\bf k}\alpha}(\tau_1)a_{-{\bf k}'\beta}(\tau_2)\rangle$ are matrices in the spin space, $\hat v_j^t$ is transposed matrix of velocity. The arguments of the zero field Green functions are denoted as $
K_{\pm}=\left(\Omega_m\pm{\omega_n}/{2}, {\bf k}\pm{\bf q}/{2}\right)$. The Matsubara frequencies 
take the values $\Omega_m=\pi (2m+1-n)T$ and $\omega_n=2\pi nT $.  
The explicit form of velocity matrix (\ref{v}) is
\begin{equation}
\hat v_j=v_j\hat\sigma_0+\hat w_j+\hat u_j,
\label{velo}
\end{equation}
where for the simplest case of isotropic medium 
\begin{equation}
v_j=k_j/m,~~~~\hat w_j=\gamma\hat\sigma_j,~~~~\hat u_j=i\frac{c}{e}\mu_Be_{lmj}\hat\sigma_l q_m,
\label{vel}
\end{equation}
and $\sigma_0$ is the unit matrix.

To make calculations, it is more convenient to pass from the spin to the band representation, where the one-particle Hamiltonian  has a diagonal form. 
In the absence of a magnetic field, it is
\begin{equation}
H_0=\sum_{\bf k}\xi_{\alpha\beta}({\bf k})a^{\dagger}_{{\bf k}\alpha}a_{{\bf k}\beta}=
\sum_{{\bf k},\lambda=\pm}\xi_{\lambda}({\bf k})c^{\dagger}_{{\bf k}\lambda}c_{{\bf k}\lambda}.
\label{Ham}
\end{equation}
 Here, the band energies are 
\begin{equation}
\xi_{\lambda}({\bf k})=\varepsilon({\bf k})-\mu+\lambda
|\mbox{\boldmath$\gamma$}({\bf k})|,
\end{equation}
such that the two Fermi surfaces are determined by the equations $\xi_{\lambda}({\bf k})=0 $. For the simplest case of a quadratic energy spectrum
in a cubic crystal,  both Fermi surfaces keep the spherical form.
The difference of the band energies $2|\mbox{\boldmath$\gamma$}({\bf k}_F)|$ characterizes   the intensity of the spin-orbital coupling. 

The Fermi momentum with $\mbox{\boldmath$\gamma$}=0$ is determined by the equation $\varepsilon({\bf k}_F)=\varepsilon_F$.
The corresponding density of states at the Fermi energy per one spin projection is $N_0=mk_F/2\pi^2$.
The split bands Fermi momenta are
\begin{equation}
k_{\pm}=k_F\mp m\gamma.
\end{equation}
Here and in all  the following calculations,  we assume  $\gamma k_F\ll\varepsilon_F$.

The diagonalization is made by the following transformation
\begin{equation}
\label{band transform}
    a_{{\bf k}\alpha}=\sum_{\lambda=\pm}u_{\alpha\lambda}({\bf k})c_{{\bf k}\lambda},
\end{equation}
with the coefficients
\begin{equation}
\label{Rashba_spinors}
   u_{\uparrow\lambda}{(\bf k})=
    \sqrt{\frac{|\mbox{\boldmath$\gamma$}|+\lambda\gamma_z}{2|\mbox{\boldmath$\gamma$}|}},~~~
    u_{\downarrow\lambda}({\bf k})=\lambda
    \frac{\gamma_x+i\gamma_y}{\sqrt{2|\mbox{\boldmath$\gamma$}|(|\mbox{\boldmath$\gamma$}|+\lambda\gamma_z)}},  
\end{equation}
forming a unitary matrix $\hat u({\bf k})$.

The zero field Green functions in the band representation are diagonal and have the following form
\cite{Sam07}:
\begin{eqnarray}
G^{(0)}_{\lambda\lambda'}(\omega_n,{\bf k})=\delta_{\lambda\lambda'}G_{\lambda}(\omega_n,{\bf k}),\nonumber\\
F^{(0)}_{\lambda\lambda'}(\omega_n,{\bf k})=\delta_{\lambda\lambda'}t_\lambda({\bf k})F_{\lambda}(\omega_n,{\bf k}),
\label{Gr}
\end{eqnarray} 
where
\begin{eqnarray}
G_{\lambda}(\omega_n,{\bf k})=-\frac{i\omega_n+\xi_\lambda}{\omega_n^2+\xi_\lambda^2+
|\tilde\Delta_{\lambda}({\bf k})|^2},\nonumber\\
F_{\lambda}(\omega_n,{\bf k})=\frac{\tilde\Delta_{\lambda}({\bf k})}{\omega_n^2+\xi_\lambda^2+
|\tilde\Delta_{\lambda}({\bf k})|^2},
\end{eqnarray} 
and
\begin{equation}
    t_{\lambda}({\bf k})=-\lambda
    \frac{\gamma_x({\bf k})-i\gamma_y({\bf k})}{\sqrt{\gamma_x^2({\bf k})+\gamma_y^2({\bf k})}}.
\end{equation}
The functions $\tilde\Delta_{\lambda}({\bf k})$ are the gaps in the $\lambda$-band quasiparticle spectrum in the superconducting state.
In the simplest model with BCS pairing interaction $v_g({\bf k},{\bf k}')= -V_g$, the gap functions are the same in both bands: $\tilde\Delta_{+}({\bf k})=\tilde\Delta_{-}({\bf k})=\Delta$ and we deal with pure singlet pairing \cite{SamMin08}.

 If in passing to the band representation
we  neglect \cite{Mineev}  the difference between $\hat u({\bf k})$ and $\hat u({\bf k}\pm{\bf q}/2)$, the expression for the current keeps the same form (\ref{basic})  with  diagonal Green function matrices (\ref{Gr}) and nondiagonal  velocity matrices obtained from Eqs. (\ref{velo}),
(\ref{vel}) by the substitutions
$\mbox{\boldmath$\hat\sigma$}\to\mbox{\boldmath$\hat \tau$}({\bf k})=\hat u^{\dagger}({\bf k})\mbox{\boldmath$\hat\sigma$}\hat u({\bf k})$ and $\mbox{\boldmath$\hat\sigma$}^t\to\mbox{\boldmath$\hat \tau$}^t({\bf -k})$:
\begin{equation}
v_j=k_j/m,~~~~\hat w_j=\gamma\hat\tau_j,~~~~\hat u_j=i\frac{c}{e}\mu_Be_{lmj}\hat\tau_l q_m,
\label{vel2}
\end{equation}

The expressions for the $\mbox{\boldmath$\hat \tau$}({\bf k})$ matrices  are
\begin{widetext}
\begin{equation}
\label{m_i}
 \hat \tau_x=
    \left(\begin{array}{cc}
      \hat\gamma_x & -
      \frac{\gamma_x\hat\gamma_z+i\gamma_y}{\gamma_\perp} \\
      -
      \frac{\gamma_x\hat\gamma_z-i\gamma_y}{\gamma_\perp} & -\hat\gamma_x \\
    \end{array}\right),~~
    \hat \tau_y=
    \left(\begin{array}{cc}
      \hat\gamma_y & -
      \frac{\gamma_y\hat\gamma_z-i\gamma_x}{\gamma_\perp} \\
      -
      \frac{\gamma_y\hat\gamma_z+i\gamma_x}{\gamma_\perp} & -\hat\gamma_y \\
    \end{array}\right),\quad
    \hat \tau_z=
    \left(\begin{array}{cc}
      \hat\gamma_z & \frac{\gamma_\perp}{\gamma} \\
      \frac{\gamma_\perp}{\gamma} & -\hat\gamma_z \\
    \end{array}\right),
\end{equation}
\end{widetext}
where $\hat{\mbox{\boldmath$\gamma$}}=\mbox{\boldmath$\gamma$}/|\mbox{\boldmath$\gamma$}|$,
$\gamma_\perp=\sqrt{\gamma_x^2+\gamma_y^2}$.  They obey  the following identities\cite{Sam07}:
\begin{eqnarray}
\tau_{i,\lambda\lambda^\prime}^t(-{\bf k})=-t^\star_\lambda t_{\lambda^\prime}\tau_{i,\lambda\lambda^\prime}({\bf k}),\nonumber\\
\tau_{i,++}\tau_{j,++}=\tau_{i,--}\tau_{j,--}=\hat\gamma_i\hat\gamma_j,\nonumber\\
\tau_{i,+-}\tau_{j,-+}=\delta_{ij}-\hat\gamma_i\hat\gamma_j+ie_{ijk}\hat\gamma_k.
\end{eqnarray}
The explicit form of the integrand in Eq. (\ref{basic}) in band representation  is
\begin{widetext}
 \begin{equation}
 \label{trace}
{\text Tr} \{  \hat v_i({\bf k})\hat G^{(0)}(K_+)\hat v_j({\bf k})\hat G^{(0)}(K_-)- \hat v_i({\bf k})\hat F^{(0)}(K_+)\hat v_j^t(-{\bf k})\hat F^{+(0)}(K_-)\}={\text Tr}_{dia}+ {\text Tr}_{para}+ {\text Tr} _{gyro}
 \end{equation}
 This formula contains rich information. The terms
 \begin{eqnarray}
{\text Tr}_{dia}=v_{i}G_+v_{j}G_++v_{i}F_+v_{j}F^\dagger_++
w_{++,i}G_+w_{++,j}G_++w_{++,i}F_+w_{++,j}F_+^{\dagger}\nonumber\\+
v_{i}G_+w_{++,j}G_++v_{i}F_+w_{++,j}F_+^{\dagger}+
w_{++,i}G_+v_{j}G_++w_{++,i}F_+v_{j}F_+^{\dagger}~+~
(+\longleftrightarrow -)
\label{dia}
 \end{eqnarray}
 determine the {\bf diamagnetic current}.  
  In the zero frequency limit,  the diamagnetic current is given by the sum of  the {\bf London current} and the {\bf current of Landau diamagnetic moment}
\begin{equation}
{\bf j}_d=-\frac{c}{4\pi\delta^2}\left ({\bf A}- \frac{ c}{2e}\nabla\varphi  \right )-c\chi_L[{\bf q}\times[{\bf q}\times{\bf  A}]].
\end{equation}

The {\bf current} corresponding to  the {\bf Pauli paramagnetic moment} 
\begin{equation}
{\bf j}_p=-c\chi_P[{\bf q}\times[{\bf q}\times{\bf  A}]]
\label{P}
\end{equation}
originates from the terms 
 \begin{eqnarray}
{\text Tr}_{para}=u_{++,i}G_+u_{++,j}G_++u_{++,i}F_+u_{++,j}F^\dagger_+\nonumber\\
+u_{+-,i}G_-u_{-+,j}G_++u_{+-,i}F_-u_{-+,j}F^\dagger_+~+~(+\longleftrightarrow -)
\label{para}
\end{eqnarray}

Finally, the {\bf gyrotropic current} originates from the terms 
 \begin{eqnarray}
{\text Tr}_{gyro}=
v_{i}G_+u_{++,j}G_++v_{i}F_+u_{++,j}F_+^{\dagger}+
u_{++,i}G_+v_{j}G_++u_{++,i}F_+v_{j}F_+^{\dagger}\nonumber\\+
u_{++,i}G_+w_{++,j}G_++u_{++,i}F_+w_{++,j}F_+^{\dagger}+w_{++,i}G_+u_{++,j}G_++w_{++,i}F_+u_{++,j}F_+^{\dagger}\nonumber\\+
u_{+-,i}G_-w_{-+,j}G_++u_{+-,i}F_-w_{-+,j}F_+^{\dagger}+
w_{+-,i}G_-u_{-+,j}G_++w_{+-,i}F_-u_{-+,j}F_+^{\dagger}
\nonumber\\+
w_{+-,i}G_-w_{-+,j}G_++w_{+-,i}F_-w_{-+,j}F_+^{\dagger}+
w_{+-,i}G_-w_{-+,j}G_++w_{+-,i}F_-w_{-+,j}F_+^{\dagger}~+~(+\longleftrightarrow -)
\label{gyro}
 \end{eqnarray}
\end{widetext}
The notation $(+\longleftrightarrow -)$ means that all these expressions are supplemented by the corresponding terms with sign $"+"$ substituted by sign  $"-"$ and vice versa.

Below we consider several properties derived from the current expression.

\section{Pauli susceptibility}

A simple example of an unusual property of non-centrosymmetric superconductors is shown by the Pauli susceptibility,   which can be found from 
the paramagnetic current, as  determined by the Eq.(\ref{para}) type terms at $\omega_n=0,~{\bf q}=0$. When the band splitting strongly exceeds the superconducting gap $\gamma k_F\gg\Delta$, the susceptibility  has a finite value  at $T=0$, even for pure s-wave pairing interaction. For isotropic superconductors it is \cite{Sam07}
\begin{equation}
\chi_P=\frac{2}{3}\mu_B^2N_0(2+Y(T)),
\label{chiP}
\end{equation}
 where 
  $$
Y(T)=-\text{T}\int_{-\infty}^{\infty}d\xi\sum_{m=-\infty}^{\infty}(G^2(\Omega_m,{\bf k})+
|F(\Omega_m,{\bf k})|^2)$$
$$=-\int_{-\infty}^{\infty}d\xi\frac{\partial f}{\partial E}
 $$
  is Yosida function (concentration of normal electrons), $f(E_{\bf k})=(e^{E_{\bf k}/T}+1)^{-1}$ is the Fermi distribution function, $E_{\bf k}=\sqrt{\xi^2+\Delta^2}$.  The corresponding paramagnetic limiting field
 \begin{equation}
 H_p=\sqrt{\frac{3}{2}}\frac{\Delta_{T=0}}{\mu _B}
 \end{equation}
is $\sqrt{ 3}$ times larger than in centrosymmetric superconductors.

Another peculiar property   is the static susceptibility anisotropy arising at finite wave vectors ${\bf q}$. This has been  theoretically predicted by Takimoto  \cite{Takimoto},  and was recently measured  by B.F\aa k at al \cite{Fak}. The analytical expression for susceptibility anisotropy
 in the case of  a  tetragonal crystal with broken space parity, point group $C_{4v}$ was  found in Ref.24. As the susceptibility at ${\bf q}=0$ the susceptibility anisotropy at finite wave vectors  can be also re-derived  from the paramagnetic current determined by the Eq.(\ref{para}) type terms at $\omega_n=0,~{\bf q}\ne 0$. The result is
\begin{equation}
 \chi_{xx}-\chi_{yy}\approx\mu_B^2N_0\gamma^2_\perp(
 q_x^2-q_y^2)(1+f({\bf q}))/\varepsilon_F^2.
 \label{ortho}
 \end{equation}
Here, the function  $f({\bf q})\sim{\cal O}\left(\frac{\gamma{\bf q}}{\varepsilon_F} \right )^2$  is  fully symmetric  with respect to the tetragonal symmetry.
Thus, the ${\bf q}$ - dependent  basal plain anisotropy proved to be quadratic on the band splitting.

The spin susceptibility in a crystal with cubic symmetry and broken space parity also loses its diagonal form:
\begin{equation}
\chi_{xy}\approx\mu^2_BN_0(i\gamma q_z/\varepsilon_F+\gamma^2q_xq_y/\varepsilon_F^2+~.~.~.).
\end{equation}

\section{Gyrotropic current}

The four lines in Eq. (\ref{gyro}) correspond to four integrals with different structure determinig gyrotropy current 
\begin{equation}
j_{gi}(\omega_n,{\bf q})
=-2e\mu_B\left [I_{1ij}+I_{2ij}+I_{3ij}   \right ]B_j+i\frac{e^2}{c}e_{ijl}{A_j}I_{4l},
\label{current}
\end{equation}
where for cubic symmetry crystal with $\hat \gamma=\hat k$
\begin{widetext}
\begin{equation}
I_{1ij}(\omega_n,{\bf q})=\int  \frac{d^3k}{(2\pi)^3}\text{T}\sum_{m=-\infty}^{\infty}v_i\hat\gamma_j
[G_+(K_+)G_+(K_-)
+F_+(K_+)F_+^\dagger(K_-)
-G_-(K_+)G_-(K_-)-F_-(K_+)F_-^\dagger(K_-)],
\end{equation}
\begin{equation}
I_{2ij}(\omega_n,{\bf q})=\gamma\int  \frac{d^3k}{(2\pi)^3}\text{T}\sum_{m=-\infty}^{\infty}\hat\gamma_i\hat\gamma_j[G_+(K_+)G_+(K_-)
+F_+(K_+)F_+^\dagger(K_-)+G_-(K_+)G_-(K_-)+F_-(K_+)F_-^\dagger(K_-)],
\end{equation}
\begin{equation}
I_{3ij}(\omega_n,{\bf q})=\gamma\int  \frac{d^3k}{(2\pi)^3}\text{T}\sum_{m=-\infty}^{\infty}(\delta_{ij}-\hat\gamma_i\hat\gamma_j)[G_+(K_+)G_-(K_-)
+F_+(K_+)F_-^\dagger(K_-)+G_-(K_+)G_+(K_-)+F_-(K_+)F_+^\dagger(K_-)],
\end{equation}
and
\begin{equation}
I_{4l}(\omega_n,{\bf q})=\gamma^2\int  \frac{d^3k}{(2\pi)^3}
\text{T}\sum_{m=-\infty}^{\infty}
\hat \gamma_l[G_+(K_+)G_-(K_-)+F_+(K_+)F^\dagger_-(K_-)
-G_-(K_+)G_+(K_-)-F_-(K_+)F^\dagger_+(K_-)].
\label{integ}
 \end{equation}
\end{widetext} 
 In all the integrals,  one has to perform summation over the Matsubara frequencies $\Omega_m$ and to pass from the discrete set of frequencies  $\omega_n$ into entire  half-plane $\omega>0$ by substitution $\omega_n\to\omega+i0$. 
 
\section{Static gyrotropy properties}
\subsection{Static gyrotropic current}
The gyrotropic current
in the static limit $\omega=0$
was first  calculated   for  an uniaxial crystal with the Rashba  spin-orbital coupling by V. Edelstein \cite{Edel}. He found that 
\begin{equation}
{\bf j}_{g}(\omega=0)= \nu (\hat c\times{\bf  B}),
\label{st}
\end{equation}
$$\nu\propto e\mu_B\gamma n_s(T)/\varepsilon_F,$$
where  $n_s(T)$ is the density of superconducting electrons, $\hat c$  is direction of anisotropy axis.

 For an isotropic medium in the static limit $\omega=0, q\to0 $,  the integral $I_{4l}$ calculated in Ref. 19 above $T_c$ vanishes. In the superconducting state it takes negligibly small value order of $\propto\Delta^2/\varepsilon^2_F$. The   other integrals  are
 \begin{equation}
  I_{1ij}=2N_0\gamma Y(T)\delta_{ij},
  \end{equation}
  \begin{equation}
  I_{2ij}=-\frac{2}{3}N_0\gamma Y(T)\delta_{ij},
  \end{equation}
  \begin{equation}
  I_{3ij}= -\frac{4}{3}N_0\gamma\delta_{ij}.
  \end{equation}
Thus, for a metal with cubic  symmetry and broken space parity, the gyrotropy current in the static limit  is 
\begin{equation}
{\bf j}_g=\nu{\bf B},~~~~\nu=\frac{8}{3}e\mu_B\gamma N_0(1-Y(T)).
\label{nu}
\end{equation}
Like the expression for the spin susceptibility  (\ref{chiP}), this formula is valid when  the band splitting strongly exceeds the superconducting gap $\gamma k_F\gg\Delta$. The expression for the static gyrotropy current free of this limitation has been found in the paper Ref. 26.

\subsection{London magnetostatics}
The generalization of the London magnetostatics to the  noncentrosymmetric case has been made by Levitov et al \cite{Levitov2}.  In this case the London current is
\begin{equation}
{\bf j}=-\frac{c}{4\pi\delta^2}\left ({\bf A}-\frac{\hbar c}{2e}\nabla\varphi\right)+\nu{\bf B},
\end{equation}
where $\delta$ is the London penetration depth and the corresponding London equation acquires the form
\begin{equation}
\Delta{\bf B}=\frac{1}{\delta^2}{\bf B}- \frac{4\pi}{c}\nu{\bf B}.
\label{London}
\end{equation}
For  a superconductor occupying the half the plane $z>0$ and the external magnetic field ${\bf H}$ directed along $x$ direction,
the boundary conditions  according to equation (7) are
\begin{equation}
{\bf B}^{int}-{\bf B}^{ext}=4\pi{\bf M}=\frac{2\pi\nu}{c}{\bf A}
\end{equation}
that is 
\begin{equation} 
B_x^{int}=H,~~~~~B_y^{int}=\frac{2\pi\nu}{c}A_y.
\label{bound}
\end{equation}
The solution of Eq. (\ref{London}) with boundary conditions (\ref{bound}) yields
\begin{eqnarray}
&&B_x+iB_y=H(1+i\tan\beta)\exp\left (-\frac{ze^{i\beta}}{\delta}\right )\nonumber\\
&&\approx H(1+i\beta)\exp\left(-\frac{i\beta z}{\delta}\right)\exp\left (-\frac{z}{\delta}\right)
\end{eqnarray}
showing that the magnetic field  inside the superconductor acquires an $y$ component, and  rotates in $x,y$ plane.
The dimensionless parameter determining these magnetostatic gyrotropy properties is an angle $\beta$:
\begin{equation}
\sin\beta=\frac{2\pi\nu\delta}{c}.
\end{equation}
A simple estimation of this angle value at $T=0$ is
\begin{equation}
\beta(T=0)=\frac{4}{3\pi}\frac{e^2}{\hbar c}\frac{\gamma}{c} k_F\delta\approx 10^{-3}.
\end{equation}
Here we used the band splitting $\gamma k_F\approx 10^3$ Kelvin (see \cite{Bose}).
This means that  London magnetostatics in noncentrosymmetric superconductors undergoes a negligibly small deviations from that  in the centrosymmetric case.

 A similar smallness is specific  for
 other gyrotropy magnetostatic properties such as the  magnetoelectric  influence on the upper critical field or the creation of helical superconducting phases. This subject was discussed in detail in Ref.29. One must make, however, an important comment. The treatment of the paper \cite{Mineev11} written properly from the symmetry point of view, has been  supported by a microscopic theory calculation. The latter was done for the single band case using the limitation
$\mu_BH<\gamma k_F$. The corresponding calculations for two bands split by  the spin-orbit coupling reproduces the same results, but all the magnetoelectric terms linear in gradients   $\eta^*K_{ij}H_iD_j\eta$ in the free energy density    acquire an additional reduction of  order $\gamma/v_F$.

\section{Dynamic gyrotropy properties}
 
 At finite frequencies, making use the relations $$E_i=\frac{i\omega}{c}A_i,~~~~~B_i=c e_{ijk}\frac{q_j}{\omega}E_k$$   one can rewrite Eq. (\ref{current}) as
 \begin{equation}
 \j_{gi}(\omega,{\bf q})=-\frac{e^2}{m\omega}[(I_{1ij}+I_{2ij}+I_{3ij})e_{jlk}q_l+me_{ilk}I_{4l}]E_k,
 \label{cur}
 \end{equation}
Here  both the frequency $\omega$ and the wave vector ${\bf q}$ dependence   of the integrals are essential. 
An important simplification
takes place in the high frequency region $\omega>qv_F$, where one can neglect  the ${\bf q}$-vector dependence of the integrals
$I_{1ij},~I_{2ij},~I_{3ij}$ and calculate $I_{4l}$ in  first order in  $q$. 
Then we have
 \begin{equation}
 I_{1ij}=0,~~~~~I_{2ij}=0,~~~~~I_{3ij}=\delta_{ij}I_3,~~~~~I_{4l}=\frac{q_l}{m}I_4,
 \label{1,2}
 \end{equation}
 \smallskip
\begin{equation}
I_3=\frac{1}{3\pi^2}\left \{k_+^2-k_-^2+a^2\ln\frac{k_+^2-a^2}{k_-^2-a^2}\right \},
\label{3}
\end{equation}
\begin{eqnarray}
I_4=-\frac{a^2}{12\pi^2}\left\{m\gamma \left (\frac{k_++m\gamma}{k_+^2-a^2}+\frac{k_--m\gamma}{k_-^2-a^2}\right )
\right.\nonumber\\
\left.+\frac{3}{2}\ln\frac{k_+^2-a^2}{k_-^2-a^2}- a^2\left ( \frac{1}{k_+^2-a^2}-\frac{1}{k_-^2-a^2} \right )\right\},
\label{4}
\end{eqnarray} 
where $a=\frac{\omega}{2\gamma}$. The integration over momenta has been performed for the case of spherical Fermi surfaces.

So, for $\omega>qv$ we arrive at the following expression for the current 
\begin{equation}
j_{gi}=ie_{ijl}\lambda(\omega)q_jE_l,~~~~~~\lambda(\omega)=\frac{ie^2}{m\omega}[I_3(\omega) +I_4(\omega)].
\label{c}
\end{equation}
As it should be \cite{electr}, $\lambda(\omega)$ is thus proved to be a pure imaginary odd function of frequency. 

Physically, the  
wave vector magnitude  is determined by the inverse skin penetration depth $q\approx\delta^{-1}$. The latter  in the infrared frequency region is of  order $10^{-5}$ cm (see the book \cite{FizKin}), such that the  integral values written above are correct for  $\omega>qv\approx10^{13}$ rad/sec. The band splitting in noncentrosymmetric metals $\sim\gamma k_F$ can be smaller or larger than this value.  In any case
at frequencies  higher than the band splitting  $\omega>\gamma k_F$, and making use 
 Eqs. (\ref{3})-(\ref{c}), we obtain a more simpler formula for gyrotropy current
\begin{equation}
j_{gi}=ie_{ijl}\lambda(\omega)q_jE_l,~~~~~~~~\lambda(\omega)=\frac{4i}{\pi^2}\frac{e^2}{\hbar}\left (\frac{\gamma k_F}{\hbar\omega}  \right )^3.
\label{43}
\end{equation}
Here we return  to dimensional units. 

The situation when  the frequency is smaller than the band splitting $\omega<\gamma k_F$ but at the same time larger than $qv_F$ has been considered in the paper\cite{Mineev},  where the integrals $I_{1ij},~I_{2ij},~I_{3ij}$ were not taken into account and
the  gyrotropy current was calculated in terms of the integral $I_4$ only.

\section{Kerr rotation}

Now we can apply the standard procedure \cite{Ben} to calculate the Kerr rotation for linearly polarized light  that is normally incident from the vacuum to the flat boundary of a medium. 
The reflected light is  elliptically polarized with the major axis rotated relative to the incident polarization by an amount 
\begin{equation}
\theta=\frac{2n\kappa\Delta \tilde n}{(1-n^2+\kappa^2)^2+(2n\kappa)^2}.
\label{e33}
\end{equation}  
Here $n$ and $\kappa$ are the real and imaginary part of  medium refraction index neglecting gyrotropy.
The  difference in the refraction indices of gyrotropy medium for circularly polarized light  with the opposite 
polarization is \cite{Mineev}
\begin{equation}
\Delta \tilde n=\tilde n_+-\tilde n_-=-\frac{4\pi}{c}\Im\lambda.
\end{equation} 
Thus,  at frequencies exceeding the band splitting,
\begin{equation}
\Delta \tilde n=-\frac{16}{\pi}\frac{e^2}{\hbar c}\left (\frac{\gamma k_F}{\hbar\omega} \right )^3. 
\end{equation}
 Although the band splitting $\gamma k_F$ is not known for many noncentrosymmetric materials, one can expect it to be about
 thousand Kelvin \cite{Bose} or in  frequency units   $\sim ~10^{14} rad/sec$. As  an example,  we  consider the 
situation where the frequency of light is of  order  this value and larger than the quasiparticle scattering rate (clean limit)
$\omega\tau\gg 1$,  and at the same time $\omega_p\gg\omega$, where $\omega_p=\sqrt{4\pi ne^2/m^*}$
is the plasma frequency.  In this frequency region, the real and imaginary part of the conductivity 
are $\sigma'\approx\omega_p^2/4\pi \omega^2\tau$ and $\sigma''\approx\omega_p^2/4\pi \omega$.
Then,  one can find $2n\kappa\approx\omega_p^2/\omega^3\tau$ and $\kappa^2-n^2\approx
\omega_p^2/\omega^2$. Thus, for the Kerr angle we obtain
\begin {equation}
\theta\approx-\frac{32}{\pi}\frac{e^2}{\hbar c}\frac{\omega}{\omega_p^2\tau}\left (\frac{\gamma k_F}{\hbar\omega} \right )^3.
\label{e3}
\end{equation}

To obtain  a numerical estimate of the Kerr angle, let us take $\omega_p\tau\approx 10^3$, $\omega/\omega_p\approx10^{-1}$, $\gamma k_F/\hbar\omega\approx1/3$. Then we find
\begin{equation}
\theta_{Kerr}\approx 1~\mu rad,
\label{Ke}
\end{equation}
which is in reasonable agreement with measured Kerr angles in the cuprate compounds reported in Refs. 4-7.

\section{Discussion}

It is interesting to compare the results obtained here with the results of calculations  performed within the model of chiral charge ordering  spinless electrons. Namely,  in the paper \cite{Hosur} the authors considered a system with Hamiltonian
\begin{equation}
H=\sum_{\bf k}E({\bf k};z)\psi^\dagger_{{\bf k},z}\psi_{{\bf k},z}-t_\perp\sum_{{\bf k},z}(\psi^\dagger_{{\bf k},z}\psi_{{\bf k},z+1}+H.c.),
\end{equation}
$$
E({\bf k};z)=\frac{1}{2m}\{{\bf k}^2+[{\bf k}\cdot{\bf n}(z)]^2    \}-E_F,
$$
$$
{\bf n}({\bf r})=n_0[\cos(\pi Qz),\sin(\pi Qz),0].
$$
Making use the following assumptions 
$$
n_0\ll1,~~~~t_\perp^2\ll n_0^2E_F,~~~~\hbar\omega\gg|t_\perp|
$$
they  have found
\begin{equation}
\lambda_{Hosur}(\omega)\approx\frac{i}{4\pi}\frac{e^2}{\hbar} \frac{n_o^4t_\perp^2E_F}{(\hbar\omega)^3},
\label{44}
\end{equation}
which is related to $\gamma(\omega)$ used in Ref.17 by  $$\gamma(\omega)=-\frac{4\pi i}{\omega}\lambda(\omega).$$

It is worth noting  that the results given by Eqs. (\ref{43}) and (\ref{44}), which originate from completely different models, have the same frequency dependence of the gyrotropy coefficient $\lambda(\omega)$.	

For completeness it should be mentioned that   two researches have recently been put forward to explain the observed Kerr effect in cuprates. \cite{Pershiguba,Aji}. Both of them  are based on the loop-current model by Varma possessing charge chirality violating space parity.

\section{Conclusion}

 The Kerr onset seen in the pseudogap phase of a large number of cuprate high-temperature superconductors, arising at about the same
 temperature as the short range charge density wave order, can serve as  evidence of a gyrotropic ordering that breaks  space inversion
symmetry but preserves time-reversal invariance. Here we proposed a simple microscopic model of an isotropic metal where inversion symmetry breaking reveals itself as a spin-orbital couling, lifting the band degeneracy and creating the electron band splitting.
The magnitude of the Kerr angle in the infrared frequency region given by Eq. (\ref{Ke})  is proved to be in reasonable agreement with recently reported observations of the Kerr effect in high T$_c$ materials.

\acknowledgments

I am quite indebted to S. Blundell for valuable help in the manuscript preparation as well
to N.N.Nikolaev who addressed me question of a witch concerning electrodynamics of noncentrosymmetric media.


\begin{thebibliography}{20}
\bibitem{fn} The corresponding expression for current  in Ref.19  contains misprint. Namely, the sign in front of the  terms with Gor'kov functions $F$ in Eq. (6) and in all following equations  of that paper should be "minus".  It has been shown in that paper that  the superconductivity introduces only small corrections $\propto\Delta^2/(\gamma k_F)^2$  to the normal state gyrotropy properties at finite frequency,  hence, the incorrect sign in front of terms determined by the Gor'kov functions does not change the main result of that paper.
\end{thebibliography}
\end{document}